# What do kets represent?

Casey Blood
CaseyBlood@gmail.com

## Abstract
It is usually assumed that a ket represents the state of an actually existing particle. But one can show there is no evidence for particles. The particle-like properties of mass, spin and charge, as well as particle-like trajectories, the photoelectric effect, and localized effects from spread-out wave functions can be explained using quantum mechanics alone. It is therefore proposed instead that kets represent particle-like solutions to a pre-representational linear partial differential equation which has Poincaré and internal symmetries. This equation underlies the completely representational character, including mass, spin, charge, internal symmetries, and symmetric and antisymmetric statistics, of current quantum mechanics.

## 1. Introduction.

This paper considers the consequences of three observations. First, it is shown in Sec. 3 that there is no evidence for objectively existing particles. Kets therefore cannot represent the states of particles so we don't currently know what they represent. Second, the same Sec. 3 arguments show that all our perceptions of a concrete, particle-like world follow from four abstract quantum mechanical principles. And third, current quantum mechanics has a completely representational structure, exactly as if it were the representation of an underlying pre-representational linear equation. In fact, it is difficult to imagine any way besides an underlying equation to explain the all-pervasive representational structure.

To give some idea of what a pre-representational equation underlying quantum mechanics might look like, we give an example in Sec. 4. It consists of a linear equation in underlying complex variables which is invariant under the Poincaré group and an SU(n) internal symmetry group. There are particle-like solutions which are functions of the underlying variables. If this were the correct theory, instead of just an example, these functions would actually *be*—not just represent—the particles. These functions, in this proposed scheme, are what the kets represent.

The underlying theory is transformed to the representational form of current quantum mechanics in several steps. (1). There are spin ½ particle-like solutions to the single variable set problem. (2). Multiple sets of variables are introduced, with an interaction term which is invariant under a permutation group. The spin ½ particle-like functions from all the different sets of variables then constitute the first set of basis vectors. (3). To accommodate antisymmetry, we allow only solutions which are antisymmetric in the sets of underlying variables. (4). Because of the antisymmetry, one can define anticommuting field operators in terms of the spin ½ basis vectors and then reformulate the theory in terms of the spin ½ field operators. (5). A Dirac-like antisymmetric vacuum composed of negative energy spin ½ functions is assumed. (6). Finally, boson states are introduced as disturbances of the vacuum. Because these are quadratic in the anticommuting fermion operators, they obey symmetric statistics.

In this way, we can construct a set of basis vectors—the vacuum, anticommuting spin ½ fermions and commuting spin 0 or 1 bosons—which gives a representational form to the theory that has the same structure as current quantum mechanics.

The initial form of the theory includes no space, time, matter or field operators. These all have to do with properties of the solutions. All physical states—the vacuum, any number of fermions, and any number of bosons are solutions of the same single equation. The arguments of Sec. 3 show that, even though the theory is abstract—all the particle-like states are functions of underlying variables rather than corresponding to actual objects—the perceptions of the observer will still match the characteristics of our familiar, concrete, particle-like world.

## 2. A problem with the ket notation

Antisymmetry exposes a problem concerning kets. To see this, suppose we have two distant spin ½ particles, with the wave function of one centered near $r_A$ and the other near $r_B$, and with both having spin $s_z = +1/2$. Then the antisymmetrized state vector of the two together is

$$|\Psi\rangle = \int d^3 r_1 \int d^3 r_2 \, \psi_{r_A}(r_1, t) \psi_{r_B}(r_2, t) \quad (1)$$
$$[|r_1, 1/2\rangle |r_2, 1/2\rangle - |r_2, 1/2\rangle |r_1, 1/2\rangle]$$

We see that the usual ket notation is insufficient here; the two kets cannot be the 'same kind' of object or else the quantity in square brackets would be zero. Actually the problem is seldom noticed because it is implicitly or explicitly assumed that the first ket in a product corresponds to the state of particle 1 and the second ket corresponds to the state of particle 2 (so in Eq. (1) the first term corresponds to particle 1 being near $r_A$, particle 2 being near $r_B$ and the second term, with the minus sign, to particle 1 being near $r_B$ and particle 2 being near $r_A$.) Thus a more complete notation for the state is

$$|\Psi\rangle = \int d^3 r_1 \int d^3 r_2 \, \psi_{r_A}(r_1, t) \psi_{r_B}(r_2, t) \quad (2)$$
$$[|r_1, 1/2\rangle_{pa\,1} |r_2, 1/2\rangle_{pa\,2} - |r_2, 1/2\rangle_{pa\,1} |r_1, 1/2\rangle_{pa\,2}]$$

In this notation, the ket in the m[th] position stands for the state of the m[th] actual, 'objectively existing' particle.

The problem is that the presumption that there is an actual particle in a definite state is, as is shown in [1] and Sec. 3, a conjecture which is impossible to defend because there is no evidence for particles. In addition, if there are actual particles, then Eq. (2) says each particle is simultaneously in two distant locations, which contradicts our usual ideas of the properties of particles. (If we assume the particle is in only one of the two possible states, then we are essentially employing a hidden variable interpretation [2,3], for which there is no convincing evidence.) Further, it seems odd to assign spin ½ to a *particle* when the quantized values of spin were derived strictly from quantum mechanics, without referring to the concept of a particle.

The situation, then, is that kets in a product must somehow be differentiated, but saying they each correspond to the state of a different particle does not work. So what we will do for the moment is simply say that each ket in a product must correspond to a different 'degree of freedom.' Eq. (1) is then written as

$$|\Psi\rangle = \int d^3r_1 \int d^3r_2\, \psi_{r_A}(r_1,t)\psi_{r_B}(r_2,t) \quad (3)$$
$$\left[|r_1, 1/2\rangle_{df\,1}|r_2, 1/2\rangle_{df\,2} - |r_1, 1/2\rangle_{df\,2}|r_2, 1/2\rangle_{df\,1}\right]$$

where *df* denotes 'degree of freedom.' The process of antisymmetrizing then corresponds to permuting the degrees of freedom. To illustrate, suppose we have a non-antisymmetrized state

$$|\Psi\rangle = \sum_{l_1,l_2} c(l_1,l_2)|l_1\rangle_{df\,1}|l_2\rangle_{df\,2} \quad (4)$$

where $l_1$ and $l_2$ denote labels. Then the (non-normalized) antisymmetrization of this state, denoted by the operator $\mathcal{A}$, yields

$$\mathcal{A}|\Psi\rangle = \sum_{l_1,l_2} c(l_1,l_2)\{|l_1\rangle_{df\,1}|l_2\rangle_{df\,2} - |l_1\rangle_{df\,2}|l_2\rangle_{df\,1}\} \quad (5)$$

where the antisymmetrizing process acts on the *df* subscripts (and no attention is paid to the order of the kets in a product).

This then is the problem; the different kets in a product must correspond to different 'degrees of freedom' but we don't currently know what the degrees of freedom correspond to.

**Speculation on what kets represent.**

In an attempt to deduce what kets represent, we presume this problem is another aspect of a general observation given in [4]. It is noted there that the current form of quantum mechanics—including the ket notation itself; the group representational origin of the particle-like properties of mass, energy, momentum, spin, and charge; antisymmetry; and the internal symmetry group representational structure of all particle states—is exactly as if current quantum mechanics is a *representational form* of a deeper, pre-representational problem. In such an approach, the kets are taken as denoting basis functions—in some currently unknown set of variables $\eta_i$—for the representation of solutions of an underlying linear equation,

$$\mathcal{O}(\eta_1, \eta_2, \ldots)\Psi(\eta_1, \eta_2, \ldots) = 0 \quad (6)$$

with solutions corresponding to physical states. The different 'degrees of freedom' of the kets then correspond to functions of different sets of variables; that is, the first ket might correspond to a function of $\eta_1$, the second to a function of $\eta_2$, and so on.

Antisymmetry is introduced by presuming the linear operator $\mathcal{O}$ is invariant under all exchanges of variables $\eta_i \leftrightarrow \eta_j$ so it is invariant under the permutation group. It is then assumed that only antisymmetric solutions are physically relevant (with the symmetric boson states arising as quadratic products of the antisymmetrized spin 1/2 states).

An example of an underlying theory is given in Sec. 4 and [4].

How could such an abstract theory, in which there are no 'objective' particles—there are only particle-*like* functions of the $\eta_i$—lead to our familiar, perceived, classical, particle-like world? We show in Sec. 3 that all the observed particle-like properties of our familiar world follow from just four basic but abstract principles, primarily linearity and relativistic invariance. So if a theory satisfies those four principles, it will give back the perceptions of our familiar, concrete world, no matter how abstract it is.

**Wigner's "unreasonable effectiveness of mathematics."**
      The idea that the various constituents of matter actually *are* functions of underlying variables (and that nothing exits besides those functions) offers a way out of the "unreasonableness" in Wigner's "unreasonable effectiveness of mathematics" in physics [5]. The conventional view is that matter is separate from the mathematics but its properties, for some unknown reason, are *described by* the solutions of certain equations. In the underlying variable approach however, the properties of matter are not just described by the solutions to an equation; instead matter *actually is* a solution to an equation [6]. In that case, the success of mathematics in describing the properties of matter is quite reasonable.

## 3. The observed properties of the physical world follow from basic quantum mechanical principles.[7]

**The basic mathematical principles of quantum mechanics are:**
(A). Linearity of the operators.
(B). Physical states correspond to vectors in a Hilbert space.
(C). Invariance and group theoretic properties.
(D). The 'local' properties of the linear, Hermitian Hamiltonian.

**The properties to be explained, without invoking collapse, hidden variables, the existence of particles, or action at a distance, are:**
(1). The perception of only one version of reality even though many can coexist in the state vector.
(2). The agreement between observers on the single version perceived.
(3). Locality: The perception of only one localized grain exposed in the double slit experiment when the wave function hits many grains; or the perceived triggering of only one localized detector in a scattering experiment when the wave function hits many detectors.
(4). The perceived particle-like trajectories observed in a bubble chamber.
(5). The particle-like properties of mass, energy, momentum, spin, and charge, along with the appropriate addition and conservation laws.
(6). The photoelectric effect.
(7). The results of Bell-like experiments on entangled systems.

**(1). The perception of only one version of reality.**
      Many versions of reality can simultaneously exist in the state vector (Schrödinger's cat is dead and alive at the same time) so we need to show this property does not conflict with our perception of a single version of reality. We state what is to be shown in a 'negative' way; there is never communicable perception of more than one version of reality in quantum mechanics.
      To illustrate we use a Stern-Gerlach experiment on a spin ½ atomic-level system S. After the magnet, the wave function corresponding to S splits into two parts, one traveling on path 1 and the other on path 2. On path j (j=1,2) we put detector Dj, and we have an observer who sees the readings of the detectors and writes the results. The state vector of the full system after S passes the magnet but before its wave function hits the detectors is then

$$|\Psi, t_1\rangle = a(1)|\text{version 1 of reality}\rangle$$
$$+a(2)|\text{version 2 of reality}\rangle \tag{7}$$

$$|\text{ver. 1}, t_1\rangle = |S1\rangle|D1, \text{no}\rangle|D2, \text{no}\rangle \ |\text{Obs sees and writes "I see no, no"}\rangle$$
$$|\text{ver. 2}, t_1\rangle = |S2\rangle|D1, \text{no}\rangle|D2, \text{no}\rangle \ |\text{Obs sees and writes "I see no, no"}\rangle$$

Note that no basis has been chosen in writing this equation. The magnet-'particle' Hamiltonian splits the wave function in this fashion no matter what basis is chosen.

Now we look at the system after S has hit the detectors on the two paths but before the observer looks at the readings. Using the time translation operator $U$ to advance the system from $t_1$ to $t_2$, we have

$$|\Psi, t_2\rangle = U(t_2, t_1)|\Psi, t_1\rangle \tag{8}$$
$$= a(1)U(t_2, t_1)|\text{ver. 1}, t_1\rangle + a(2)U(t_{,2}, t_1)|\text{ver. 2}, t_1\rangle$$

But the effect of the linear operator $U(t_2,t_1)$ on $|\text{ver. 1}, t_1\rangle$ is independent of a(1) and a(2), so it can be calculated at a(1)=1, a(2)=0. This is equivalent to saying branch 1 evolves as if branch 2 is not there (a(2)=0). Generalizing, we have the result;

**All branches evolve in time as if the other branches were not there. Thus no photons, no signal, no information of any kind, can be passed from one branch to another. The branches are totally isolated from each other so they constitute separate universes.**

So at time $t_2$, we have

$$|\Psi, t_2\rangle = a(1)|\text{ver. 1}, t_2\rangle + a(2)|\text{ver. 2}, t_2\rangle \tag{9}$$
$$|\text{ver. 1}, t_2\rangle = |S1\rangle|D1, \text{yes}\rangle|D2, \text{no}\rangle \ |\text{Obs does not look}\rangle$$
$$|\text{ver. 2}, t_2\rangle = |S2\rangle|D1, \text{no}\rangle|D2, \text{yes}\rangle \ |\text{Obs does not look}\rangle$$

We have used the 'local' property of the interactions here; $|S1\rangle$ can only activate detector 1 and $|S2\rangle$ can only activate detector 2. Again, no basis (and in particular no 'preferred' basis) has been used to obtain this result; it follows strictly from the local nature of the Hamiltonian.

Finally we let the system evolve to time $t_3$, after the observer has looked. Then we get

$$|\Psi, t_3\rangle = U(t_3, t_2)|\Psi, t_2\rangle$$
$$= a(1)U(t_3, t_2)|\text{ver. 1}, t_2\rangle + a(2)U(t_{,3}, t_2)|\text{ver. 2}, t_2\rangle$$

But $U(t_3, t_2)|\text{ver. 1}, t_2\rangle$ is independent of $a(1)$ and $a(2)$ so we can set $a(1)=1$, $a(2)=0$ (as if version 2 is simply not there). In that case, since the version of the observer can perceive only what occurs on her own branch, we obtain

$$U(t_3, t_2)|\text{ver. 1}, t_2\rangle = |S, 1\rangle|D1, \text{yes}\rangle|D2, \text{no}\rangle$$
$$|\text{Ver. 1 of obs sees only yes, no and writes "I see only yes, no."}\rangle$$
$$= |\text{Ver. 1}, t_3\rangle$$

with a similar result for version 2. We therefore have

$$|\Psi, t_3\rangle = a(1)|\text{ver. 1}, t_3\rangle + a(2)|\text{ver. 2}, t_3\rangle$$
$$|\text{ver. 1}, t_3\rangle = |S, 1\rangle|D1, \text{yes}\rangle|D2, \text{no}\rangle \qquad (10)$$
$$\qquad |\text{Ver. 1 of obs sees only yes, no and writes "I see only yes, no."}\rangle$$
$$|\text{ver. 2}, t_3\rangle = |S, 2\rangle|D1, \text{no}\rangle|D2, \text{yes}\rangle$$
$$\qquad |\text{Ver. 2 of obs sees only no, yes and writes "I see only no, yes."}\rangle$$

As in the $t_1$ and $t_2$ cases, no basis was assumed in arriving at this $t_3$ result; it is dictated by the detector-photon and photon-observer interactions.

This argument shows that "I see something other than a single, 'everyday' version of reality" is never written in quantum mechanics. Thus *the laws of quantum mechanics can never lead to a situation in which more than one version of reality is communicably perceived.* Also, note that the versions perceived correspond to eigenvectors of the relevant apparatus.

(A note on 'communicably perceived.' The point is to show that no-particle quantum mechanics agrees with our *perceptions*. And that implicitly means perceptions we can tell others about. We will concede that the arguments do not apply to non-communicable perceptions, whatever these might be, but they are irrelevant to the point at hand.)

**(2). No disagreement among observers.**

If we have two observers, A and B, then there are still only two branches. And the versions of A and B on each branch can perceive only what occurs on that branch. Thus the versions on each branch must agree with each other so there is never disagreement among observers in quantum mechanics.

**(3). Localized perceived effect from a spread-out wave function.**

Suppose we have a scattering experiment in which the scattered wave function for the 'particle' S spreads out and hits every one of N detectors $|Dj\rangle$. Just before the wave function reaches the detectors, we divide it into the sum of N separate terms, $|S, j\rangle$, with the $j^{\text{th}}$ term about to hit the $j^{\text{th}}$ detector, so the total wave function is

$$|\Psi\text{ before}\rangle = \sum_{j=1}^{n} a(j)|S, j\rangle \prod_{k=1}^{N} |Dk, no\rangle \qquad (11)$$

When the wave function hits the detectors, the $|S, j\rangle$ term changes only detector j from no to yes, so the state vector becomes the sum of N terms, with only one detector reading yes in each term.

$$|\Psi\text{ after}\rangle = \sum_{j=1}^{N} a(j)|S, j\rangle|D1, \text{no}\rangle|D2, \text{no}\rangle \ldots |Dj, \text{yes}\rangle \ldots |DN, \text{no}\rangle \qquad (12)$$

(Note: This same state vector can be arrived at in a somewhat more satisfactory way by integrating the equation of continuity and changing the volume integral over $\nabla \cdot J$ to a sum of N area integrals. The a(j) is then related to the 'amount' of the wave function that passed through detector j.)

And now we can revert to the argument in **(1)**. Each of these N terms constitutes a separate branch and so the version of the observer on that branch will perceive only what occurs on that branch, namely that one and only one detector reads yes, while all the others read no. That is, one obtains a perceived *localized* outcome—only one localized detector activated—in spite of the wave function hitting all the detectors! (But quantum mechanics doesn't tell us which detector "I" will perceive as activated.)

This perceived localization effect is one of the main reasons for assuming particles exist. So its explanation in terms of quantum mechanics alone is a severe blow to the idea of particles.

**(4). Particle-like trajectories.**

This is just **(3)** applied multiple times, with each potential nucleation center treated as a detector. Note that the quantum mechanics leads to the perception of 'continuous,' not disjoint trajectories.

**(5). The particle-like properties of mass, energy, momentum, spin and charge.**

These, along with locality, have been the defining characteristics of particles since 1700 and before. But one can show, using group representation theory, that they are actually properties of state vectors (whether particles exist or not). First, experimental results do not depend on *where* an experiment is done, or *when*. Further, Einstein deduced relativistic invariance—that the equations of physics must be invariant under space-time 'rotations' that leave $x^2+y^2+z^2-t^2$ invariant. When combined with linearity, these invariances have several consequences. The invariance under *where* and *when* implies the solutions—the state vectors—can be labeled by energy and momentum. Further these mathematical labels correspond exactly to the physically measurable properties of energy and momentum. And one gets conservation of energy and momentum; for an isolated system, the total energy and momentum remain the same forever.

Second, if we also take into account invariance under Einstein's relativistic rotations, so we get the full Poincaré group, then the solutions acquire two additional properties—spin and mass. The allowed values for spin are integer multiples of 1/2; 0, 1/2, 1, 3/2, 2,…. These two results are astonishing! Mass itself, perhaps the most basic property of matter, is a consequence of the simple observation that the outcomes of experiments don't depend on the orientation (generalized to include relativistic rotations) or position of the apparatus! And the mathematically allowed quantized values for the angular momentum—and *only* these values—are exactly observed; electrons, quarks and neutrinos, for example, have spin ½ while the photon has spin 1 (and no particle has, for example, spin 2/5).

The last basic property of matter is charge. The charges have the same invariance-under-rotation type of origin as spin, only in a somewhat peculiar way. It was found experimentally that the proton, neutron and all other strongly interacting 'elementary' particles were made up of three more elementary particles—quarks. The only way to make sense of all the various experiments was to suppose the quarks came in three 'colors' (nothing to do with actual colors, of course; just colorful language). Further experiment showed that the theory had to be invariant under rotations in the color space, just as the hydrogen atom theory is invariant under rotations in our ordinary three-dimensional space. And the group theoretic *labels* associated with this invariance are the strong charges.

A similar argument can also be made for the electromagnetic and weak charges, with all charges quantized as integer multiples of the basic charges. (In the case of quarks, the basic electrical charge is 1/3 the charge on the electron.) Thus, just as mass and so on are labels, and physical properties, associated with the wave function due to invariance under space-time transformations, so the three types of charges are labels, and physical properties, associated with the wave function due to an internal—"within the particle," so to speak; not in extended space and time—set of rotations in some abstract space.

We also note that if there are several 'particles' present, group representation theory correctly predicts that the energies, momenta, z components of spin, and charges all add algebraically. It also implies the conservation laws for these quantities. And it correctly predicts the somewhat more complicated addition rules for total spin.

## 6. The photoelectric effect and Compton scattering.

The photoelectric effect was the experiment which prompted Einstein to conclude that light was indeed a particle, and the Compton scattering of light off electrons seemed to confirm this idea. The thinking in the photoelectric effect was that the incoming light wave was so spread out that each individual electron could not receive enough energy to (quickly) knock it out of a metal crystal. But if one assumed there was a particle hidden within the wave, and used classical conservation of energy and momentum, one could explain the results very nicely. The same particle explanation also gave a correct answer for the related problem of Compton scattering of light (x-rays) off electrons.

But it turns out these two experimental results can be explained without resorting to the hypothesis of particles. In quantum mechanics, energy and momentum belong to the state vector, and the usual addition and conservation rules apply, so particles are not needed on this account. Further one can show that the classical idea of a spread-out wave transferring only a small amount of energy to a small 'target' does not hold in quantum mechanics. Instead each small 'piece' of light carries, and can transfer, the full amount of energy and momentum to an electron (basically because the linear equations of motion are independent of the coefficients on the wave functions, so the small coefficient when a spread-out wave hits an electron does not inhibit the transfer of a large energy and momentum to the electron; see [1]). Thus particles are not necessary to explain the photoelectric and Compton effects.

## 7. The Bell-Aspect experiment.

There are a number of experiments involving entangled wave functions—for example the Bell-Aspect experiment [8,9], Wheeler's delayed-choice experiment [10], and the quantum eraser [11]—where the results are difficult to understand if one assumes the physical world is made of particles, but they follow simply from no-particle quantum mechanics. To illustrate, we will summarize the Bell-Aspect experiment, which is a descendant of the arguments of Einstein, Podolsky, and Rosen [12].

Two photons are (nearly) simultaneously emitted from an atom which both before and after the emission is in a spin 0 state. One photon travels to the left and the other to the right. The right beam of photons is split into a + polarization and a – polarization by an apparatus oriented at angle 0 and the left beam is split into a + polarization and a – polarization by an apparatus oriented at an angle $\theta$. For each pair of photon wave functions that go through the apparatus, there are four possible outcomes. The probabilities for each of these predicted by quantum mechanics are

$$
\begin{aligned}
&+L\text{:}+R \quad sin^2\theta/2 \\
&+L\text{:}-R \quad cos^2\theta/2 \\
&-L\text{:}+R \quad cos^2\theta/2 \\
&-L\text{:}-R \quad sin^2\theta/2
\end{aligned} \quad (13)
$$

These probabilities are experimentally confirmed in the Aspect experiment.

But now suppose we consider the experiment from the classical particle point of view. Then there would be one photon moving to the left in a definite polarization state and another moving to the right, also in a definite polarization state. It was Bell's stroke of genius to find a way to show that the four probabilities in Eq. (13) could not hold in the classical picture in which each of the two localized photon particles possesses a definite state of polarization that is not changed by a measurement on the other, distant photon.

Suppose, however, that one insists on the classical picture of localized carriers of the particle-like properties. Then to account for Eq. (13) holding experimentally, one must postulate that there is some unknown force which instantaneously changes the polarization state of the second, distant particle when the first is measured. But it is not necessary to postulate such a peculiar force if one assumes only the state vectors exist; quantum mechanics perfectly predicts the results of Eq. (13) without the presumption of an instantaneous-action-at-a-distance force. The long-range correlations (which are *not* interactions) between the two photon states implied by Eq. (13) are built into the entangled wave function.

The same conclusion holds for all entangled-state experiments. Quantum mechanics alone, with no postulated action at a distance, correctly and simply predicts the results. But if one assumes there are particles in a definite, localized state, one must twist and turn to find an explanation.

**No evidence for particles. Wave-particle duality.**

For the idea of actual, 'objective existing' particles to be viable, there must be some experiment that requires particles for its explanation. But we see that all the experiments which *seem* to require particles can instead be explained by the properties of the wave functions/state vectors alone. Thus there is no evidence for particles. Wave-particle duality is not a duality in the actual nature of matter; instead it is only a dichotomy in the properties—wave-like (interference) *and* particle-like (mass, spin, charge, perceived localization)—of the state vectors.

## 4. Example of an underlying pre-representational theory.

We have seen in Sec. 3 that the general mathematical principles (A) – (D) ensure that quantum mechanics yields a theory which agrees with the general nature of our perceptions—only one version of reality perceived, locality, the particle-like properties of mass, spin, charge and so on. It doesn't matter how abstract the scheme is, the theory will still give back a perceptually valid model of the physical world. We have also noted (Sec. 2 and [4]) that current quantum mechanics is structured entirely—the use of kets, representations of the Poincaré and internal symmetry group representation theory, antisymmetry—as if it were the representation of an underlying, pre-representational theory. To illustrate what a pre-representational theory might look like, we will give an example here which takes the form of a linear, partial differential equation in some set of variables, $\eta_i$ as in Eq. (6) ( see also [4,13-16]). It satisfies principles (A) (linearity), (B) (a Hilbert space), and (C) (invariance principles). But we don't see how to check principle (D) on the local Hamiltonian, so it cannot be considered at this time as a candidate for *the* underlying theory. (See Ch. 4 of [17] for a possible way to proceed in attempting to verify (D). There, a property of the interaction Hamiltonian in quantum field theory is shown to imply cluster decomposition.)

### A. The relevant groups and the underlying variables.

First we have the homogeneous Lorentz group, $G^1{}_4$, which is the set of all four-dimensional 'rotations' of x,y,z,t that leave $x^2+y^2+z^2-t^2$ invariant. In terms of the usual x,y,z,t variables (which will not be used here) its six generators are

$$J_1 = -i\left(y\frac{\partial}{\partial z} - z\frac{\partial}{\partial y}\right), J_2 = -i\left(z\frac{\partial}{\partial x} - x\frac{\partial}{\partial z}\right), J_3 = -i\left(x\frac{\partial}{\partial y} - y\frac{\partial}{\partial x}\right)$$

$$K_1 = -it\frac{\partial}{\partial x} - ix\frac{\partial}{\partial t}, K_2 = -it\frac{\partial}{\partial y} - iy\frac{\partial}{\partial t}, K_3 = -it\frac{\partial}{\partial z} - iz\frac{\partial}{\partial t} \qquad (14)$$

with these satisfying the commutation relations

$$[J_i, J_j] = i\varepsilon_{ijk}J_k, [J_i, K_j] = i\varepsilon_{ijk}K_k, [K_i, K_j] = -i\varepsilon_{ijk}J_k \qquad (15)$$

If we now include the four space-time translations, with generators

$$P_0 = H = \frac{i\partial}{\partial t}, \qquad P_j = -\frac{i\partial}{\partial x_j} \qquad (16)$$

they commute with one another,

$$[P_\mu, P_\nu] = 0 \qquad (17)$$

and satisfy

$$[J_i, P_j] = i\varepsilon_{ijk}P_k, \ [K_i, P_j] = -iP_0, [K_i, P_0] = -iP_i, \ [J_i, P_0] = 0 \qquad (18)$$

Curiously, the set of all complex rotations with determinant 1 of the two complex variables u and v is homomorphic (with a 2→1 map) to the homogeneous Lorentz group. Its generators, which obey the same commutation relations as in Eq. (15), are

$$\begin{aligned}
J_1 &= \frac{1}{2}\left(u_{bi}\frac{\partial}{\partial v_{bi}} + v_{bi}\frac{\partial}{\partial u_{bi}} - \bar{u}_{bi}\frac{\partial}{\partial \bar{v}_{bi}} - \bar{v}_{bi}\frac{\partial}{\partial \bar{u}_{bi}}\right) \\
J_2 &= \frac{i}{2}\left(-u_{bi}\frac{\partial}{\partial v_{bi}} + v_{bi}\frac{\partial}{\partial u_{bi}} - \bar{u}_{bi}\frac{\partial}{\partial \bar{v}_{bi}} + \bar{v}_{bi}\frac{\partial}{\partial \bar{u}_{bi}}\right) \\
J_3 &= \frac{1}{2}\left(u_{bi}\frac{\partial}{\partial u_{bi}} - v_{bi}\frac{\partial}{\partial v_{bi}} - u_{bi}\frac{\partial}{\partial u_{bi}} + v_{bi}\frac{\partial}{\partial v_{bi}}\right) \\
K_1 &= \frac{i}{2}\left(u_{bi}\frac{\partial}{\partial v_{bi}} + v_{bi}\frac{\partial}{\partial u_{bi}} + \bar{u}_{bi}\frac{\partial}{\partial \bar{v}_{bi}} + \bar{v}_{bi}\frac{\partial}{\partial \bar{u}_{bi}}\right) \\
K_2 &= \frac{-1}{2}\left(-u_{bi}\frac{\partial}{\partial v_{bi}} + v_{bi}\frac{\partial}{\partial u_{bi}} + \bar{u}_{bi}\frac{\partial}{\partial \bar{v}_{bi}} - \bar{v}_{bi}\frac{\partial}{\partial \bar{u}_{bi}}\right) \\
K_3 &= \frac{i}{2}\left(u_{bi}\frac{\partial}{\partial u_{bi}} - v_{bi}\frac{\partial}{\partial v_{bi}} + u_{bi}\frac{\partial}{\partial u_{bi}} - v_{bi}\frac{\partial}{\partial v_{bi}}\right)
\end{aligned} \qquad (19)$$

The *b* and *i* subscripts are summed over, with *b* running from 1 to 2 and *i* from 1 to n, with n equal, say, to 6. These have to do with an assumed SU(n) internal symmetry group. The transformation properties of the various quantities in terms of the basis vectors $|j\rangle$ and $|\bar{j}\rangle$ for the *n* and $\bar{n}$ representations (which are inequivalent) are

$$|j\rangle: u_{1j}, v_{1j}, \partial \bar{u}_{1j}, \partial \bar{v}_{1j}, \bar{u}_{2j}, \bar{v}_{2j}, \partial u_{2j}, \partial v_{2j} \tag{20}$$
$$|\bar{j}\rangle: u_{2j}, v_{2j}, \partial \bar{u}_{2j}, \partial \bar{v}_{2j}, \bar{u}_{1j}, \bar{v}_{1j}, \partial u_{1j}, \partial v_{1j}$$

Because $|j\rangle|\bar{j}\rangle$ (summed from j=1 to n) is invariant under transformations from SU(n), we see that the J,K generators of SL(2), Eq. (19), are invariant under SU(n).

Sets of the 4n complex variables $u_{bi}$, $v_{bi}$ will be the underlying variables in our example. Space-time variables $x_\mu$ do not enter into the pre-representational form of the theory. They are added later with the aid of the $P_\mu$ translation operators.

**B. The linear operator.**

The linear operator for this 'single-particle' example is chosen to be a harmonic oscillator-like partial differential operator (because it leads to a solvable equation) in the u,v;

$$\mathcal{O} = -\left(\frac{\partial}{\partial u_{1i}}\frac{\partial}{\partial v_{2i}} - \frac{\partial}{\partial v_{1i}}\frac{\partial}{\partial u_{2i}} + \frac{\partial}{\partial \bar{u}_{1i}}\frac{\partial}{\partial \bar{v}_{2i}} - \frac{\partial}{\partial \bar{v}_{1i}}\frac{\partial}{\partial \bar{u}_{2i}}\right) \tag{21}$$
$$+ (u_{1i}v_{2i} - v_{1i}u_{2i} + \bar{u}_{1i}\bar{v}_{2i} - \bar{v}_{1i}\bar{u}_{2i})$$

This has a large invariance group. Among the generators that commute with it are the J, K of Eq. (19). The four operators

$$P_0 = -u_{1i}\frac{\partial}{\partial \bar{v}_{2i}} + v_{1i}\frac{\partial}{\partial \bar{u}_{2i}} + \bar{u}_{1i}\frac{\partial}{\partial v_{2i}} - \bar{v}_{1i}\frac{\partial}{\partial u_{2i}}$$
$$P_1 = -u_{1i}\frac{\partial}{\partial \bar{u}_{2i}} + v_{1i}\frac{\partial}{\partial \bar{v}_{2i}} + \bar{u}_{1i}\frac{\partial}{\partial u_{2i}} - \bar{v}_{1i}\frac{\partial}{\partial v_{2i}}$$
$$P_2 = i\left(+u_{1i}\frac{\partial}{\partial \bar{u}_{2i}} + v_{1i}\frac{\partial}{\partial \bar{v}_{2i}} + \bar{u}_{1i}\frac{\partial}{\partial u_{2i}} + \bar{v}_{1i}\frac{\partial}{\partial v_{2i}}\right) \tag{22}$$
$$P_3 = u_{1i}\frac{\partial}{\partial \bar{v}_{2i}} + v_{1i}\frac{\partial}{\partial \bar{u}_{2i}} - \bar{u}_{1i}\frac{\partial}{\partial v_{2i}} - \bar{v}_{bi}\frac{\partial}{\partial u_{2i}}$$

also commute with $\mathcal{O}$. The 10 generators of Eqs. (19) and (22) obey the commutation relations of the Poincaré group, and so $\mathcal{O}$ is relativistically invariant. We see from Eq. (20) that $\mathcal{O}$, as well as the ten generators, are also invariant under SU(n).

**C. The scalar product and basis vectors.**

If we are to have a Hilbert space, we need a scalar product in the space of functions of the u,v. We choose

$$\langle f|g\rangle = \prod_{b,i} \int d\Re(u_{bi})d\Im(u_{bi})d\Re(v_{bi})d\Im(v_{bi}) \tag{23}$$
$$\bar{f}(u_{bi}, v_{bi}, \bar{u}_{bi}, \bar{v}_{bi})g(u_{bi}, v_{bi}, \bar{u}_{bi}, \bar{v}_{bi})$$

where $\Re, \Im$ stand for the real and imaginary parts resp. One can show that this scalar product is invariant under the Poincaré group defined by the generators of Eqs. (19) and (22) as well as under SU(n).

There are solutions to the equation $\mathcal{O}\,\psi = \lambda\,\psi$ that are eigenvectors of the four momenta operators. Curiously, the normalization is not the usual $\delta^3(p-p')$ but instead is $\delta^4(p-p')$. The reason is that there is a generator G of the invariance group which obeys $[G,P_\mu]=i\,\lambda' P_\mu$ and so the mass of solutions scales continuously.

### D. Interactions.

We now suppose there are N sets of 4n complex variables

$$\{u_{bi}^{(m)}, v_{bi}^{(m)}\} = \eta^{(m)}, \quad m = 1, \ldots, N;\ b = 1,2;\ i = 1, \ldots, n \tag{24}$$

We can then put in an interaction by choosing the linear operator to be

$$\mathcal{O} = \sum_{m=1}^{N} \mathcal{O}^{(1)}(\eta^{(m)}) + \sum_{m' \neq m} \mathcal{O}^{(2)}(\eta^{(m')}, \eta^{(m)}) \tag{25}$$

where $\mathcal{O}^{(1)}$ is the 'single-particle' operator of Eq. (21) and the *form* of the interaction term $\mathcal{O}^{(2)}$ is independent of $m', m$. The interaction term is assumed to be invariant under the full Poincaré group, which implies it must commute with the total translation operators

$$P_\mu^{(T)} = \sum_{m=1}^{N} P_\mu^{(m)} \tag{26}$$

with $P_\mu^{(m)}$ defined in Eq. (22). Thus we must have

$$\left(P_\mu^{(m)} + P_\mu^{(m')}\right) \mathcal{O}^{(2)}\left(\eta^{(m)}, \eta^{(m')}\right) = 0 \tag{27}$$

with $P_\mu^{(m)} \mathcal{O}^{(2)} \neq 0$ so there is a non-zero interaction. One possible way to satisfy this equation is to suppose there are functions $x_\mu(\eta^{(m)}) \equiv x_\mu^{(m)}$ such that

$$\left[P_0^{(m)}, x_\nu^{(m)}\right] = +i\delta_{0\nu},\ \left[P_j^{(m)}, x_0^{(m)}\right] = 0,\ \left[P_j^{(m)}, x_k^{(m)}\right] = -i\delta_{jk} \tag{28}$$

and then set

$$\mathcal{O}^{(2)}(\eta^{(m)}, \eta^{(m')}) = \mathcal{O}^{(2)}\left(x_\mu^{(m)} - x_\mu^{(m')}\right) \tag{29}$$

where, dropping the superscripts,

$$x_\mu = c_{kk'} z_{\mu,kk'}/I \tag{30}$$

with the $c_{kk'}$ real, $c_{k'k} = -c_{kk'}$ and

$$I = c_{kk\prime}(u_{1k}v_{1k\prime} - v_{1k}u_{1k\prime}) + cc$$
$$z_{0,kk\prime} = u_{2k}\bar{u}_{1k\prime} + v_{2k}\bar{v}_{1k\prime} + cc$$
$$z_{3,kk\prime} = u_{2k}\bar{u}_{1k\prime} - v_{2k}\bar{v}_{1k\prime} + cc \tag{31}$$
$$z_{1,kk\prime} = u_{k2}\bar{v}_{k\prime 1} + v_{k2}\bar{u}_{k\prime 1} + cc$$
$$z_{2,kk\prime} = i(-u_{k2}\bar{v}_{k\prime 1} + v_{k2}\bar{u}_{k\prime 1}) + cc$$

I is an invariant under SL(2) and the $z_\mu$ transform like a four-vector. We can make $\mathcal{O}^{(2)}$ an ISL(2) invariant if we make it a function of

$$(x_\mu^{(m)} - x_\mu^{(m\prime)})(x_\mu^{(m)} - x_\mu^{(m\prime)}) \tag{32}$$

The operator will still have an internal symmetry, but it will be a subset of SU(n) and will depend on the values of the $c_{kk\prime}$.

This is just an example of how one might introduce an interaction. There are many other possibilities. When one finds the correct interaction term and converts it to representational form in terms of fields, it will presumably be of the quantum field theoretic $J^{t,\mu}A^{t,\mu}$ form required by gauge invariance, where *t* runs over the generators of the invariance group.

**E. Antisymmetry.**

We see that the linear operator of Eq. (25) is invariant under exchanges of variable sets $\eta^{(m)} \leftrightarrow \eta^{(m\prime)}$ and so it is invariant under the permutation group of N objects. We assume the only representation of physical interest is the totally antisymmetric one so all states will be antisymmetrized under the exchange of variables.

**F. Spin ½ states as basis vectors. The vacuum**.

The conversion to the current representational form of the theory is done in two steps. First we use basis vectors in each set of $\eta^{(m)}$ that are particle-like (i.e., they have energy and momentum) spin ½ solutions of the 'single-particle' equation $\mathcal{O}^{(1)}\psi = \lambda\psi$ with $\mathcal{O}^{(1)}$ given in Eq. (21). The negative energy spin ½ states are used to form a vacuum state which, for now, is supposed to be Dirac-like, constructed from negative energy basic spin ½ states, and completely antisymmetrized. It is not difficult to construct vacuum states which are either completely invariant under the internal symmetry group or have a partially broken symmetry. One can also arrange the structure of the vacuum so that a gauge transformation—an an SU(n) transformation in which the matrix elements vary with position—of the vacuum induces a term proportional to the derivative of the gauge transformation.

Positive energy spin ½ states, fermions, will be, roughly, basic positive energy spin ½ states superimposed on the vacuum, with the whole state still being antisymmetric. Antiparticles will correspond to holes (missing states) in the vacuum.

Because of the antisymmetry, one can convert from the ket representational form of the theory, where each ket stands for a function of a particular $\eta^{(m)}$, to a quantum field (QFT) representational form, in which the theory is expressed entirely in terms of the vacuum state and anticommuting spin1/2 creation and annihilation operators. The $\mathcal{O}^{(2)}$ of Eq. (25) in this representational form will have a four-fermion form.

There are interpretations/understandings of quantum mechanics in which the field operators are considered to be the most basic constituents. That is not true here; instead the field operators arise from a combination of the functional basis vectors and antisymmetry, with the functional basis vectors being the most basic constituents.

### G. Interaction-mediating boson states.

The second step in the conversion to the representational form of current quantum mechanics is to introduce interaction-carrying bosons—analogous to, or in the correct theory identical to gluons, photons, the Z and the Ws. These are presumed to result from the positive energy spin ½ states interacting with the negative energy spin ½ states in the vacuum to give 'disturbances' of the vacuum. To illustrate, suppose the four-fermion interaction is of the form $J^{t,\mu}J'^{t,\mu}$ with both $J$s quadratic in the fermion field operators and the $t$ indicating an operator that transforms like the $t^{\text{th}}$ generator of SU(n). We then suppose that the $J'^{t,\mu}$ acting on the vacuum gives a 'disturbance' of the vacuum which is to be equated with the interaction-mediating boson field $A^{t,\mu}$. Because $J'^{t,\mu}$ is quadratic in the fermion fields, $A^{t,\mu}$ will obey symmetric statistics.

The antisymmetric and symmetric statistics then allow one to set up the whole theory in terms of anticommuting fermion and commuting boson field operators, and the interaction Hamiltonian will then have the field-theoretic boson-mediated $J^{t,\mu}A^{t,\mu}$ form. The internal-symmetry structure of the vacuum—unbroken or broken—may affect the form of the interaction as well as the purely bosonic terms.

And to reiterate the main point, every physically relevant state, including the vacuum and states with any number of fermions and bosons, corresponds to a solution of the single, linear, pre-representational equation (which includes no mention of space, time, or matter).

### H. Time evolution.

A solution of the linear equation contains the state vector for all time [6]. However, the time translation operator of Eq. (22) shifts the whole solution in time and is not of interest in what we think of as the evolution in time. Instead, one must do the following: Using the space-time translation generators of Eq. (22), one can overlay a space-time grid on the $\eta^{(m)}$ space (see [16]). The time evolution of interest is then how the solution varies from one equal-time surface to another. This will be deducible from the linear operator.

### I. Gravity.

Is it possible to include gravity in this approach? One might try the following [16]. Assume a macroscopic concentration of matter affects the vacuum state, perhaps by slightly (G is small) changing the local energy-momentum density in the vacuum on a macroscopic scale. This, then, would affect the flight of free particles and thereby cause a gravity-like effect. To match this to general relativity, one would have to show that concentrations of matter do indeed change the macroscopic properties of the vacuum, and then relate those changes to changes in the space-time grid.

## Summary

We deduce from antisymmetry that each ket in a product corresponds to a different 'degree of freedom.' It is normally assumed the degrees of freedom refer to the states of objectively existing particles. But it is shown in Sec. 3 that there is no evidence for particles. All the reputed evidence, including the photoelectric effect, localized effects from spread-out wave functions, particle-like trajectories and so on can be explained by using the mathematics of quantum mechanics alone. Thus some other, non-particle way of accounting for the degrees of freedom must be found.

A clue to the nature of these degrees of freedom is to observe that mass, energy, momentum, spin, charge, internal symmetries, symmetric and antisymmetric statistics, and even kets themselves are all representational constructs. That is, the structure of quantum mechanics is exactly as if it is the *representational form* of a pre-representational linear equation.

This suggests the following: Underlying the current representational form of quantum mechanics is a pre-representational, linear partial differential equation in some unknown set of variables. This equation is presumed to be invariant under a group of transformations of the variables which includes the Poincaré group, internal symmetries, and exchanges of variable sets. Antisymmetrized functions of the underlying variables with all the properties of spin ½ particles are used as basis vectors to recover the current representational form of quantum mechanics. It is these functions that correspond to the ket degrees of freedom. Interaction-carrying bosons are disturbances of the vacuum which are quadratic in the spin ½ states. All states, including the vacuum and any number of fermions and bosons, correspond to—or rather, actually are—functions of the underlying variables, with the functions all satisfying the same, single underlying equation.

An example of an underlying pre-representational equation is given in which the variables are complex, with no mention of space-time or particles. In spite of its abstract nature—nothing exists besides these functions of the underlying complex variables—it can be shown that such a theory can lead to the perceptions associated with our familiar, concrete, particle-like world.

It is our opinion that the completely representational structure of current quantum mechanics can only be accounted for by assuming there is an underlying pre-representational equation. But a great deal of work remains to be done. The major tasks are that we need guiding principles to help deduce the proper form of the linear operator. And we need ways, probably using the ideas of gauge invariance, to deduce the correct form for the vacuum. We also need to find the connection with the field theoretic variational principle.

# References:

[1] Casey Blood, "No evidence for particles", *arXiv*:quant-ph/0807.3930v2 (2011).
[2] David Bohm, "A suggested interpretation of quantum theory in terms of "hidden" variables", *Phys. Rev.* **85** 166,180 (1952).
[3] D. Bohm and B. J. Hiley *The Undivided Universe* (Routledge, New York, 1993).
[4] Casey Blood, "Group Representational Clues to a Theory Underlying Quantum Mechanics", *arXiv*:quant-ph/0903.3160, (2009).
[5] E. P. Wigner, "The unreasonable effectiveness of mathematics in the natural sciences", *Comm. On Pure and Applied Mathematics*, **13**: 1-14 (1960).
[6] Max Tegmark, "The mathematical universe". *Foundations of physics*, **38**, 101-150 (2007).
[7] The line of reasoning which emphasizes how most of our perceptions could be accounted for by the mathematics of quantum mechanics alone was initiated by Hugh Everett III in "Relative State Formulation of Quantum Mechanics", *Rev. Mod. Phys.* **29**, 454-462 (1957).
[8] J. S. Bell, "On the Einstein Podolsky Rosen paradox", *Physics*, **1**, 195 (1964).
[9] A. Aspect, P. Grangier, and G. Rogers, "Experimental Realization of Einstein-

Podolsky-Rosen-Bohm Gedankenexperiment: A New Violation of Bell's Inequalities", *Phys. Rev. Lett*. **47**, 460 (1981).
[10] V. Jacques *et al*, "Experimental Realization of Wheeler's delayed-choice Gedankenexperiment", *Science* **315**, 966 (2007).
[11] S. P. Walborn, M. O. Terra Cunha, S. Paua, and C. H. Monken, "Double Slit Quantum Eraser", *Phys. Rev. A*, **65** 033818, (2002).
[12] Einstein, A, Podolsky, B., Rosen, N. "Can quantum-mechanical description of physical reality be considered complete?" *Phys. Rev*. **47** (10): 777–80, (1935).
[13] F. A. Blood, Jr. "Independent variables in quantum mechanics". *J. Math. Phys*. **22**, 67-77, (1981).
[14] F. A. Blood, "A relativistic quantum mechanical harmonic oscillator without space-time variables". *J. Math Phys.* **29**, 1389-1395 (1988).
[15] F. A. Blood, "Interaction-mediating vector bosons as collective oscillations of the vacuum"*, Il Nuovo Cimento* 102, 1059-1081 (1989).
[16] Casey Blood, "A linear theory underlying quantum mechanics", *arXiv*:quant-ph/1211.7337 (2012).
[17] Steven Weinberg, *The quantum theory of fields, vol. I*, Cambridge University Press, (1995).